# Effective de Sitter-Schwarzschild Metrics

Moorad Alexanian

*Department of Physics and Physical Oceanography*
*University of North Carolina Wilmington, Wilmington, NC 28403-5606*

E-mail: alexanian@uncw.edu

**Abstract.** An earlier work that replaced the de Sitter-Schwarzschild metric for a cosmological constant $\Lambda \leq 0$ by an effective anti-de Sitter black hole analytic at $\Lambda = 0$ is here extended to the horizon structure through its three possible roots. We replace the de Sitter-Schwarzschild metric for all possible values of $\Lambda$ and the Schwarzschild radius $R$ for the three possible available roots by either a Schwarzschild black hole whose radius depends on both $R$ and $\Lambda$ or a de Sitter space whose cosmological constant depends on both $R$ and $\Lambda$.

## 1. Introduction

The de Sitter-Schwarzschild metric, also known as the Kottler metric, is an exact solution to Einstein's field equations in general relativity that describes the spacetime geometry surrounding a spherically symmetric, non-rotating mass $M$ in the presence of a positive cosmological constant $\Lambda > 0$. It represents a black hole embedded within de Sitter space, which models an exponentially expanding universe driven by a cosmological constant, and features both a black hole event horizon and a cosmological horizon beyond which the expansion dominates. This metric generalizes the Schwarzschild solution for an isolated black hole in asymptotically flat spacetime by incorporating the effects of cosmic expansion, providing a simple yet fundamental model for black holes in universes with dark energy-like components [1].

The statistical entropy of a Schwarzschild-anti-de Sitter black hole was determined [2] by approximating the de Sitter-Schwarzschild metric with that of the Schwarzschild metric by the root that is analytic in the cosmological constant at $\Lambda = 0$. The latter is one of the three possible singularities of the de Sitter-Schwarzschild metric. In addition, the root was only expressed as a power series expansion in terms of the cosmological constant $\Lambda$. In the present work, we still approximate the de Sitter-Schwarzschild metric by the Schwarzschild metric but, instead, consider the two positive roots individually, however, we do so with the exact value given rise to a black hole event horizon, which depends on the value of the cosmological constant $\Lambda$ and the Schwarzschild radius $R$. In addition, we also approximate the de Sitter-Schwarzschild metric individually by a de Sitter singularity for the three real roots, which gives rise, instead, to a cosmological de Sitter horizon, which depends on the Schwarzschild radius $R$ and the cosmological constant $\Lambda$.

This paper is arranged as follows. In Sec. 1, we post a rather brief introduction to the de Sitter-Schwarzschild metric and present our effective de Sitter and effective Schwarzschild metrics obtained from the de Sitter-Schwarzschild metric. In Sec. 2, we use the roots of the de Sitter-Schwarzschild metric to generate our effective metrics for each of the possible roots determined by $f(r) = 0$, which gives us a black hole event horizon and a cosmological horizon. Finally, Sec. 3 summarizes our results.

## 2. De Sitter–Schwarzschild metric

In the de Sitter-Schwarzschild metric, where the line element takes the static, spherically symmetric form in Schwarzschild-like coordinates:

$$ds^2 = -\left(1 - \frac{R}{r} - \frac{\Lambda r^2}{3}\right)dt^2 + \left(1 - \frac{R}{r} - \frac{\Lambda r^2}{3}\right)^{-1} dr^2 + r^2\left(d\theta^2 + \sin^2\theta d\varphi^2\right), \tag{1}$$

where the speed of light in vacuum $c = 1$, the Schwarzschild radius $R = 2MG$, with $M$ the mass of the black hole, $G$ the Newtonian constant of gravitation, and the cosmological constant $\Lambda$. The metric (1) reduces to the Schwarzschild metric when $\Lambda = 0$ and to the de Sitter metric when $R = 0$ and $\Lambda > 0$, indicating the interpolation between these two limiting cases. The de Sitter-Schwarzschild metric with a negative cosmological constant ($\Lambda < 0$) admits black hole solutions which are asymptotic to anti-de Sitter space [3, 4].

The de Sitter-Schwarzschild metric has singularities associated with the roots of $f(r)$ that determine the horizon structure through its three possible real roots

$$f(r) = 1 - \frac{R}{r} - \frac{\Lambda r^2}{3}, \tag{2}$$

which are

$$r_0(\Lambda, R) = \frac{1}{2}\frac{2^{1/3}\left(2^{1/3}\left(\left(-3R + \sqrt{\frac{9\Lambda R^2 - 4}{\Lambda}}\right)\Lambda^2\right)^{2/3} + 2\Lambda\right)}{\Lambda\left(\left(-3R + \sqrt{\frac{9\Lambda R^2 - 4}{\Lambda}}\right)\Lambda^2\right)^{1/3}}, \tag{3}$$

and

$$r_\pm(\Lambda, R) = -\frac{1}{2\Lambda}\left(\Lambda r_0(\Lambda, R) \mp \sqrt{-3\Lambda^2 r_0^2(\Lambda, R) + 12\Lambda}\right). \tag{4}$$

The root $r_0(\Lambda, R)$ is real, positive for $-\infty < \Lambda \leq 4/(9R^2)$, complex for $\Lambda > 4/(9R^2)$, preventing a naked singularity, and discontinuous at $\Lambda = 0$. The root $r_+(\Lambda, R)$ is real, positive for $0 \leq \Lambda \leq 4/(9R^2)$ and complex elsewhere, preventing a naked singularity. The root $r_-(\Lambda, R)$ is real, negative (unphysical) for $\Lambda > 0$ and complex elsewhere. The effective, positive Schwarzschild radius $r(\Lambda, R)$ extends the traditional Schwarzschild radius $R$ by incorporating dependence on the cosmological constant for the range $-\infty < \Lambda \leq 4/(9R^2)$.

Fig. 1 (a)-(d) show the effective Schwarzschild radius $r(\Lambda, R)$ vs. $\Lambda$ that characterizes the de Sitter-Schwarzschild metric by an effective black hole that follows from the positive roots of (2), viz., $r_0(\Lambda, R)$ and $r_+(\Lambda, R)$, as a function of $\Lambda$ for given value of $R$ and for $-\infty < \Lambda \leq 4/(9R^2)$. For $0 < \Lambda \leq 4/(9R^2)$, the existence of two radii for a given value of $\Lambda$ is reminiscent of a first-order phase transition with $\Lambda = 4/(9R^2)$ as the critical point. The critical point corresponds to the Nariai limit [1], the occurrence of a double zero, where the black hole event horizon and the cosmological horizon coincide, viz., $r_0(\Lambda_c, R) = r_+(\Lambda_c, R)$, where $\Lambda_c = 4/(9R^2)$ and $r_c(\Lambda_c, R) = 1/\sqrt{\Lambda_c} = 3R/2$.

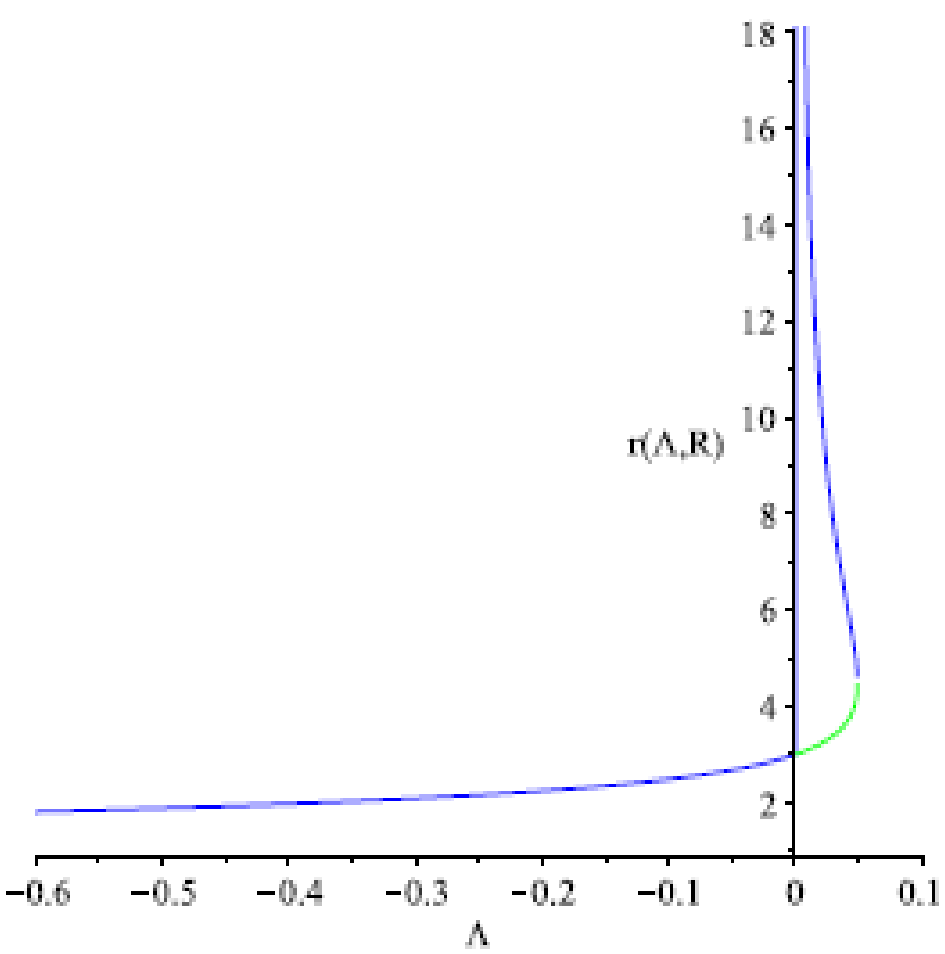

(a) Fig. 1. Plot of effective Schwarzschild radius $r(\Lambda, R)$ vs. $\Lambda$, where $r(\Lambda, R) = r_0(\Lambda, 3)$ (blue) and $r(\Lambda, R) = r_+(\Lambda, 3)$ (green). Critical point $\Lambda_c(3) = 4/81 = 0.049...$

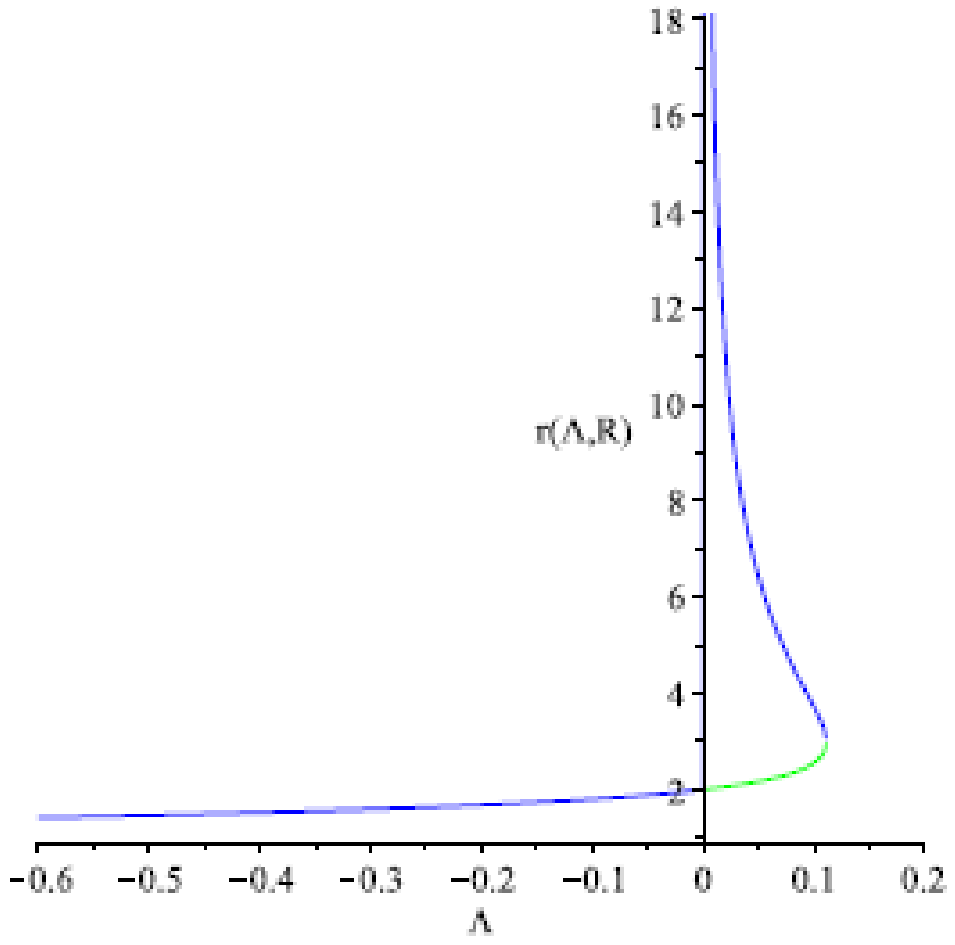

(b) Fig. 1. Plot of effective Schwarzschild radius $r(\Lambda, R)$ vs. $\Lambda$, where $r(\Lambda, R) = r_0(\Lambda, 2)$ (blue) and $r(\Lambda, \mathrm{R}) = r_+(\Lambda, 2)$ (green). Critical point $\Lambda_c(2) = 1/9 = 0.111....$

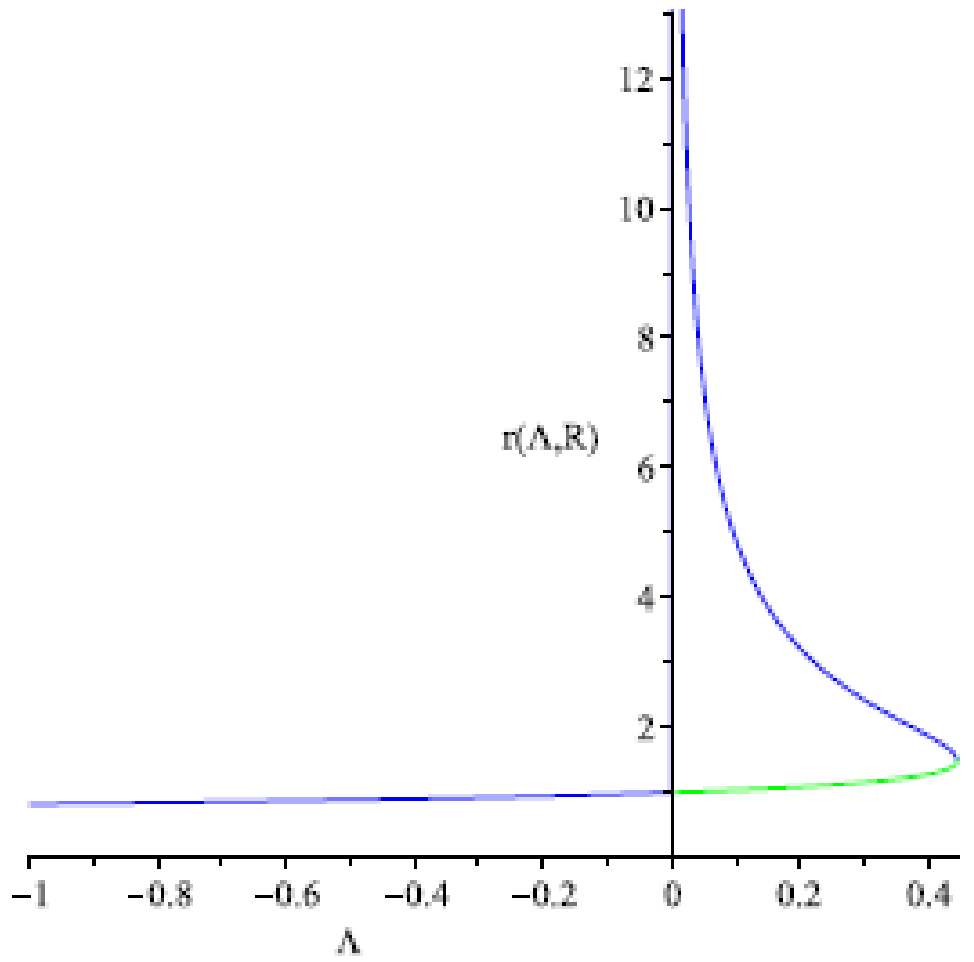

(c) Fig. 1. Plot of effective Schwarzschild radius $r(\Lambda, R)$ vs. $\Lambda$, where $r(\Lambda, R) = r_0(\Lambda, 1)$ (blue) and $r(\Lambda, R) = r_+(\Lambda, 1)$ (green). Critical point $\Lambda_c(1) = 4/9 = 0.444....$

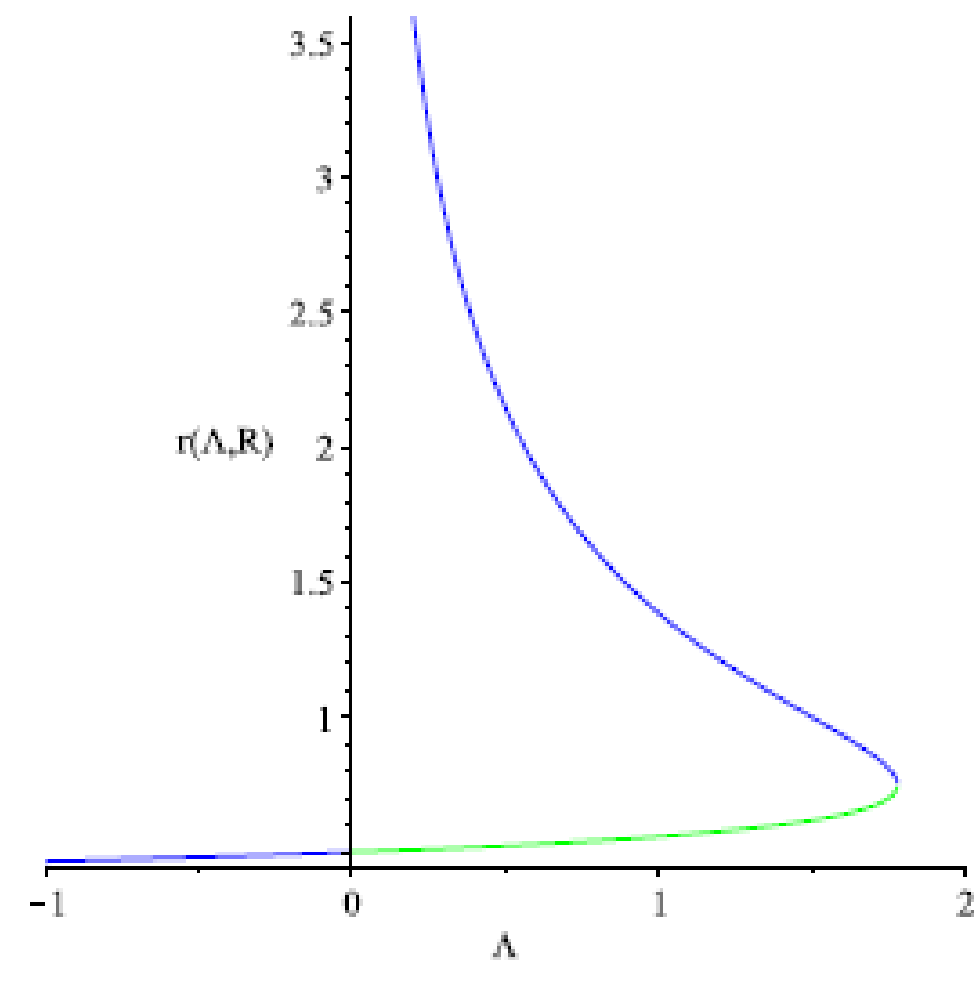

(d) Fig. 1. Plot of effective Schwarzschild radius $r(\Lambda, R)$ vs. $\Lambda$, where $r(\Lambda, R) = r_0(\Lambda, 0.5)$ (blue) and $r(\Lambda, R) = r_+(\Lambda, 0.5)$ (green). Critical point $\Lambda_c(0.5) = 16/9 = 1.778...$

Fig. 2 (a)-(d) show the effective cosmological constant $\lambda(\Lambda, R)$ vs. $\Lambda$ that characterizes the de Sitter-Schwarzschild metric by an effective de Sitter metric that follows from the three roots of (2) given by (3) and (4), viz., $\lambda(\Lambda, R) = 3/r_i^2(\Lambda, R)$, where $i = 0, \pm$. We now have three roots rather than two roots as in Fig. 1 (a)-(d) for $0 < \Lambda \leq 4/(9R^2)$. Hence, for a given value of $\Lambda$, one has a sort of triple point, where three phases coexist. The critical point corresponds to the Nariai limit [1], the occurrence of a double zero, where the black hole event horizon and the cosmological horizon coincide, viz., $r_0(\Lambda_c, R) = r_+(\Lambda_c, R)$, where $\Lambda_c = 4/(9R^2)$ and $\lambda_c(\Lambda_c, R) = 3\Lambda_c = 4/(3R^2)$. Note that $\partial\lambda(\Lambda, R)/\partial R < 0$.

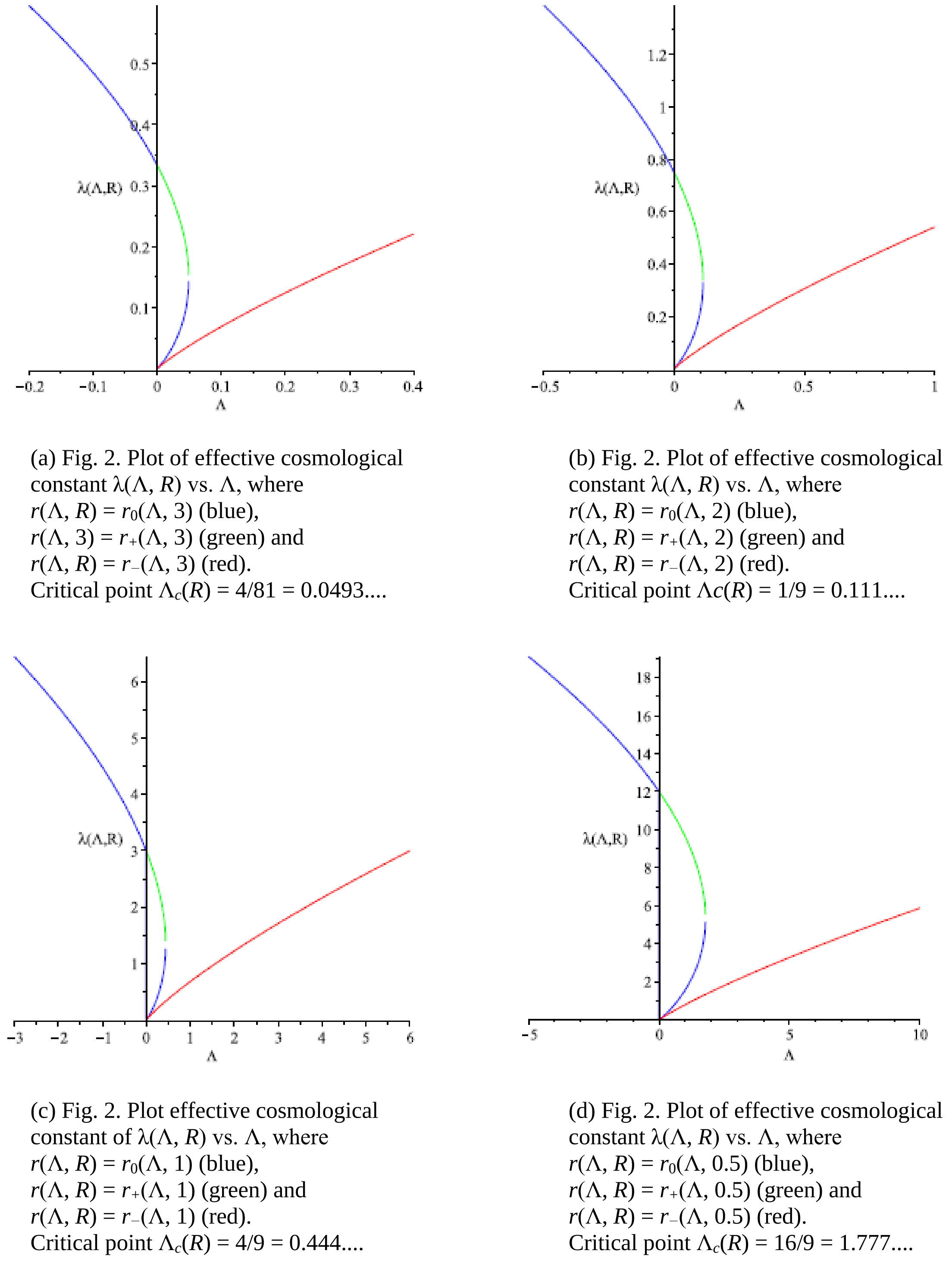

(a) Fig. 2. Plot of effective cosmological constant $\lambda(\Lambda, R)$ vs. $\Lambda$, where
$r(\Lambda, R) = r_0(\Lambda, 3)$ (blue),
$r(\Lambda, 3) = r_+(\Lambda, 3)$ (green) and
$r(\Lambda, R) = r_-(\Lambda, 3)$ (red).
Critical point $\Lambda_c(R) = 4/81 = 0.0493$....

(b) Fig. 2. Plot of effective cosmological constant $\lambda(\Lambda, R)$ vs. $\Lambda$, where
$r(\Lambda, R) = r_0(\Lambda, 2)$ (blue),
$r(\Lambda, R) = r_+(\Lambda, 2)$ (green) and
$r(\Lambda, R) = r_-(\Lambda, 2)$ (red).
Critical point $\Lambda c(R) = 1/9 = 0.111$....

(c) Fig. 2. Plot effective cosmological constant of $\lambda(\Lambda, R)$ vs. $\Lambda$, where
$r(\Lambda, R) = r_0(\Lambda, 1)$ (blue),
$r(\Lambda, R) = r_+(\Lambda, 1)$ (green) and
$r(\Lambda, R) = r_-(\Lambda, 1)$ (red).
Critical point $\Lambda_c(R) = 4/9 = 0.444$....

(d) Fig. 2. Plot of effective cosmological constant $\lambda(\Lambda, R)$ vs. $\Lambda$, where
$r(\Lambda, R) = r_0(\Lambda, 0.5)$ (blue),
$r(\Lambda, R) = r_+(\Lambda, 0.5)$ (green) and
$r(\Lambda, R) = r_-(\Lambda, 0.5)$ (red).
Critical point $\Lambda_c(R) = 16/9 = 1.777$....

Fig. 3 (a)-(b) show the effective Schwarzschild radius $r(\Lambda, R)$ vs. $R$ that characterizes the de Sitter-Schwarzschild metric by an effective black hole that follows from the positive roots of (2), viz., $r_0(\Lambda, R)$ and $r_+(\Lambda, R)$, as a function of $R$ for given value of $\Lambda$ and for $-\infty < \Lambda \leq 4/(9R^2)$. For $0 < \Lambda \leq 4/(9R^2)$, the existence of two radii for a given value of $R$ is reminiscent of a first-order phase transition with $\Lambda = 4/(9R^2)$ at the critical point, where $\Lambda_c = 4/(9R^2)$ and $r_c(\Lambda_c, R) = 1/\sqrt{\Lambda_c}$. The critical point corresponds to the Nariai limit [1], the occurrence of a double zero, where the black hole event horizon and the cosmological horizon coincide, viz., $r_0(\Lambda_c, R) = r_+(\Lambda_c, R)$, where $\Lambda_c(\Lambda, R) = 4/(9R^2) = 3\Lambda_c$ and $r_c(\Lambda, R) = 1/\sqrt{\Lambda}$. Fig. 3 (c)-(d) show the effective cosmological constant $\lambda(\Lambda, R) = 3/r_i^2(\Lambda, R)$, where $i = 0, \pm$, as a function of $R$ and $\lambda_c(\Lambda, R) = 3\Lambda$.

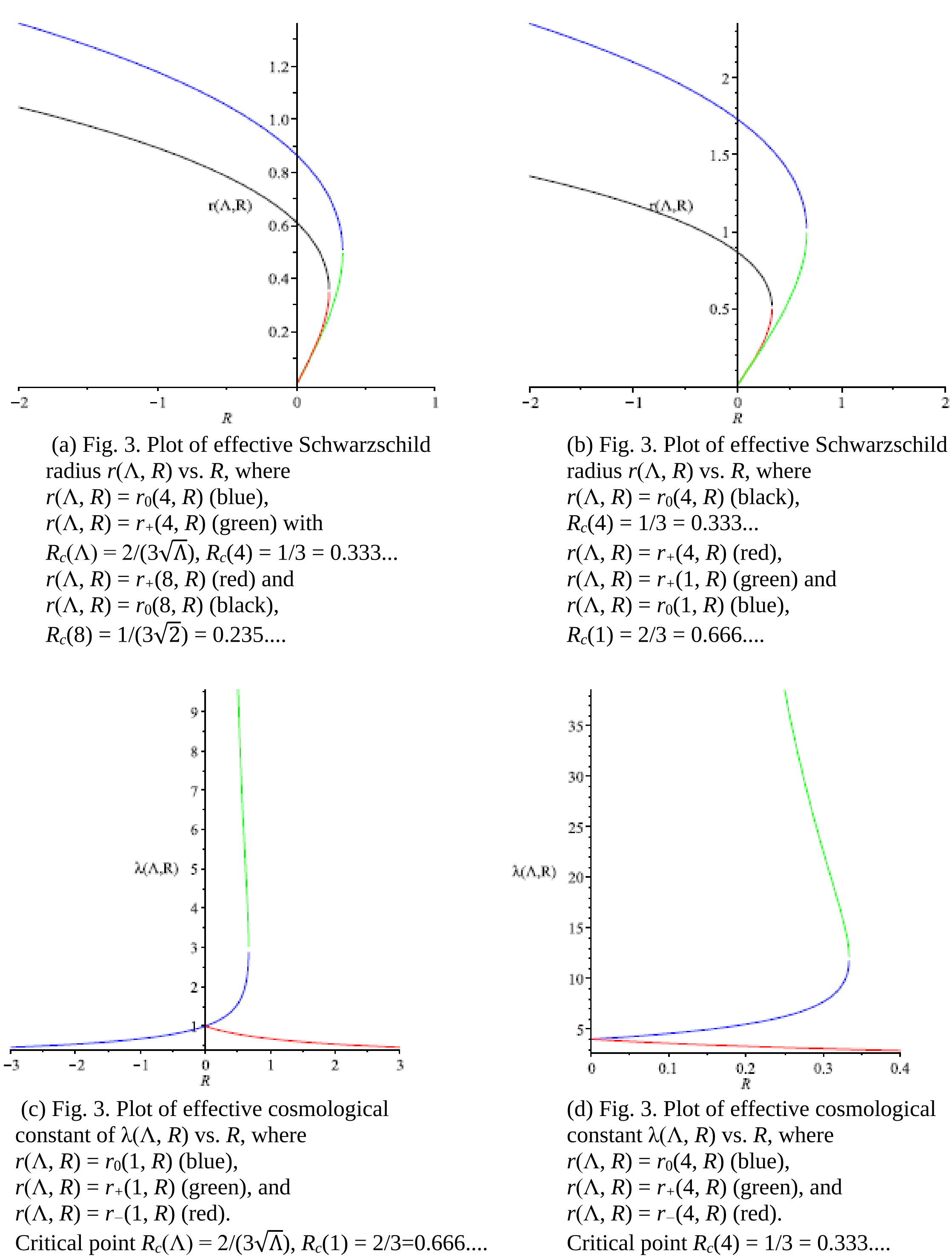

(a) Fig. 3. Plot of effective Schwarzschild radius $r(\Lambda, R)$ vs. $R$, where $r(\Lambda, R) = r_0(4, R)$ (blue), $r(\Lambda, R) = r_+(4, R)$ (green) with $R_c(\Lambda) = 2/(3\sqrt{\Lambda})$, $R_c(4) = 1/3 = 0.333...$ $r(\Lambda, R) = r_+(8, R)$ (red) and $r(\Lambda, R) = r_0(8, R)$ (black), $R_c(8) = 1/(3\sqrt{2}) = 0.235....$

(b) Fig. 3. Plot of effective Schwarzschild radius $r(\Lambda, R)$ vs. $R$, where $r(\Lambda, R) = r_0(4, R)$ (black), $R_c(4) = 1/3 = 0.333...$ $r(\Lambda, R) = r_+(4, R)$ (red), $r(\Lambda, R) = r_+(1, R)$ (green) and $r(\Lambda, R) = r_0(1, R)$ (blue), $R_c(1) = 2/3 = 0.666....$

(c) Fig. 3. Plot of effective cosmological constant $\lambda(\Lambda, R)$ vs. $R$, where $r(\Lambda, R) = r_0(1, R)$ (blue), $r(\Lambda, R) = r_+(1, R)$ (green), and $r(\Lambda, R) = r_-(1, R)$ (red). Critical point $R_c(\Lambda) = 2/(3\sqrt{\Lambda})$, $R_c(1) = 2/3=0.666....$

(d) Fig. 3. Plot of effective cosmological constant $\lambda(\Lambda, R)$ vs. $R$, where $r(\Lambda, R) = r_0(4, R)$ (blue), $r(\Lambda, R) = r_+(4, R)$ (green), and $r(\Lambda, R) = r_-(4, R)$ (red). Critical point $R_c(4) = 1/3 = 0.333....$

## 3. Conclusions

We obtain effective metrics from the de Sitter-Schwarzschild metric by reducing the metric either to a Schwarzschild solution or to a de Sitter metric. The effective metrics follow from the exact roots of the cubic equation $f(r) = 0$. This approach leads to an effective Schwarzschild radius $r(\Lambda, r)$ for the Schwarzschild metric and to an effective cosmological constant λ(Λ, *R*) for the de Sitter metric. In the former case, two of the positive roots contribute to generate two values for *r*(Λ, *r*), whereas in the latter case the three real solution generate three positive values for λ(Λ, *R*) since λ(Λ, *R*) = $3/r_i^2$(Λ, *R*), where *i* = 0, ±. In total, we generate 5 effective metrics from the original de Sitter-Schwarzschild metric, two for the effective radius *r*(Λ, *r*) of the Schwarzschild metric and three for the effective cosmological constant λ(Λ, *R*) in the de Sitter metric.

**References**

[1] De Sitter-Schwarzschild metric – Grokipedia.
[2] M. Alexanian, Arm. J. Phys. **12** (2019) 178.
[3] S.H. Hawking, D.N. Page, Commun. Math. Phys. **87** (1983) 577.
[4] H.C. Ohanian, R. Ruffin, Gravitation and Spacetime (W.W. Norton & Co., New York, 1994, 397).